\documentclass[12pt,preprint]{aastex}

\slugcomment{}

\shortauthors{Matheson et al.}
\shorttitle{Spectroscopic Variability of GRB 021004}

\begin{document}


\title{The Spectroscopic Variability of GRB 021004\altaffilmark{1}}

\author{T.~Matheson\altaffilmark{2},
P.~M.~Garnavich\altaffilmark{3},
C.~Foltz\altaffilmark{4}, 
S.~West\altaffilmark{4},
G.~Williams\altaffilmark{4},
E.~Falco\altaffilmark{5},
M.~L.~Calkins\altaffilmark{5},
F.~J.~Castander\altaffilmark{6, 7},
E.~Gawiser\altaffilmark{7, 8},
S.~Jha\altaffilmark{9},
D.~Bersier\altaffilmark{2},
and K.~Z.~Stanek\altaffilmark{2}}

\altaffiltext{1}{Based on data from the MMTO 6.5m telescope, the FLWO
1.5m telescope, and the Magellan 6.5m Baade telescope.}

\altaffiltext{2}{Harvard-Smithsonian Center for Astrophysics, 60
Garden Street, Cambridge, MA 02138; tmatheson, dbersier,
kstanek@cfa.harvard.edu.}

\altaffiltext{3}{Department of Physics, University of Notre Dame, 225
Nieuwland Science Hall, Notre Dame, IN 46556;
pgarnavi@miranda.phys.nd.edu.} 
 
\altaffiltext{4}{Multiple Mirror Telescope Observatory, University of
Arizona, 933 North Cherry Avenue, Tucson, AZ 85721;
cfoltz, swest, gwilliams@as.arizona.edu.}

\altaffiltext{5}{Smithsonian Institution, F. L. Whipple Observatory,
670 Mt. Hopkins Road, P.O. Box 97, Amado, AZ 85645; efalco,
mcalkins@cfa.harvard.edu.}

\altaffiltext{6}{Institut d'Estudis Espacials de Catalunya/CSIC, Gran
Capit\`a 2-4, 08034 Barcelona, Spain; fjc@ieec.fcr.es.}

\altaffiltext{7}{Universidad de Chile, Casilla 36-D, Santiago, Chile.}

\altaffiltext{8}{Astronomy Department, Yale University, P.O. Box
208101, New Haven, CT 06520; gawiser@astro.yale.edu.}

\altaffiltext{9}{Department of Astronomy, University of California at
Berkeley, Berkeley, CA 94720; saurabh@astron.berkeley.edu.}


\begin{abstract}
We present spectra of the optical transient (OT) associated with GRB
021004.  The spectra show a blue continuum with superposed absorption
features and one emission line.  We confirm two intervening metal-line
systems at $z = 1.380$ and $z = 1.602$ and one very strong absorption
system at a redshift of $z = 2.323$.  Ly$\alpha$ emission is also seen
at this redshift.  While the spectrum of the OT overall cannot be
simply described with a power law, the spectral index over the range
$5500-8850$ \AA\ is steep, $F_{\nu} \propto \nu^{-0.96 \pm 0.03}$.
Comparison of spectra from multiple epochs shows a distinct color
evolution with the OT becoming redder with time over the first three
days.  This is the first clear example of color change in an OT
detected spectroscopically.
\end{abstract}
\section{Introduction}

The recent rapid progress in the study of gamma-ray bursts (GRBs;
Klebesadel, Strong, \& Olson~1973) has been heralded by the swift
localization and follow-up of their afterglows at longer wavelengths
(e.g., Piro et al.~1996; Groot et al.~1997; Frail et al.~1997).
Optical spectroscopy of the afterglow of GRB~970508 (Bond~1997) led to
the first redshift determination of a GRB, with $z \geq 0.835$
(Metzger et al.~1997), settling the controversy regarding the distance
scale for these objects.

Over the last few years, optical spectroscopy of GRB afterglows has
become relatively common, and the focus of study has shifted beyond
simply constraining GRB redshifts. The afterglow spectra provide a
means to study the local environment of GRBs and connections to their
origins, for example, within massive star populations or star-forming
regions (Vreeswijk et al.~2001; Castro et al.~2002; Mirabal et
al.~2002b), as well as to study host-galaxy properties such as the
extinction law (e.g., Jha et al.~2001) and intervening material along
the line of sight.  In addition, spectropolarimetry (Barth et
al. 2002) has the potential to reveal much about the nature of GRBs
and their environments.

Combined with afterglow photometry, optical spectroscopy at later
times has provided tantalizing hints of a connection between GRBs and
core-collapse supernovae (e.g., for GRB~011121/SN~2001ke; Garnavich et
al.~2002). Furthermore, spectroscopy of GRB host galaxies (after the
afterglow has faded) also yields valuable data bearing on the question
of the origin of GRBs (e.g., Bloom, Djorgovski, \&
Kulkarni~2001). Nonetheless, early-time spectroscopic observations of
GRB afterglows (when they are still bright and easier to study!)
provide a unique back-lit view of the GRB environment and its host
galaxy.

Here we report spectroscopic observations of GRB~021004.  GRB~021004
was detected by the FREGATE, WXM, SXC instruments aboard HETE II at
12:06:13.57 UT on 2002 October~4 (Shirasaki et al.~2002). Fox et
al.~(2002a) reported their discovery of a bright, fading optical
transient (OT) within the small SXC error circle located at
$\alpha$~=~$00^{\rm h}26^{\rm m}54\fs69$,
$\delta$~=~$+18\arcdeg55\arcmin41\farcs3$ (J2000.0), and identified
this as the GRB optical afterglow. Fox et al.~(2002b) were the first
to report on optical spectroscopy of the afterglow, constraining the
GRB redshift to $z \geq 1.60$ based on the detection of two
absorption-line systems. Additional reports of spectroscopy noted
unidentified absorption lines (Eracleous et al.~2002; Sahu et
al.~2002; Castander et al.~2002), with the puzzle solved by Chornock
\& Filippenko~(2002) who identified absorption and emission lines at
$z = 2.323$ in their Keck Observatory spectroscopy of the afterglow.
The coincidence of the redshifted wavelength of Ly$\alpha$ at $z =
2.33$ and \ion{C}{4} $\lambda\lambda$1448, 1551 at $z = 1.60$ as well
as the relatively weak Ly$\alpha$ forest led to our initial mistaken
claim that $z \lesssim 1.7$ (Castander et al. 2002).  In addition,
several of the strong absorption lines consist of multiple components
(Salamanca et al. 2002; Mirabal et al. 2002a; Savaglio et al. 2002),
some with a large velocity separation interpreted as evidence for a
supernova (Salamanca et al. 2002) or high-velocity mass loss from a
Wolf-Rayet star (Mirabal et al. 2002a).

\section{Observations}
Spectra of the optical transient (OT) associated with GRB 021004 were
obtained over several nights with the 6.5m MMT telescope, the 1.5m
Tillinghast telescope at the F. L. Whipple Observatory (FLWO), and
the Magellan 6.5m Baade telescope.  The spectrographs used were the
Blue Channel (Schmidt et al. 1989) at the MMT, FAST (Fabricant et
al. 1998) at FLWO, and LDSS2 (Mulchaey 2001) at Magellan (see Table 1
for observational details).  The observations were reduced in the
standard manner with IRAF\footnote{IRAF is distributed by the National
Optical Astronomy Observatories, which are operated by the Association
of Universities for Research in Astronomy, Inc., under cooperative
agreement with the National Science Foundation.} and our own routines.
Spectra were optimally extracted (Horne 1986).  Wavelength calibration
was accomplished with HeNeAr lamps taken immediately after each OT
exposure.  Small-scale adjustments derived from night-sky lines in the
OT frames were also applied.  Spectrophotometric standards are listed
in Table 1.  We attempted to remove telluric lines using the
well-exposed continua of the spectrophotometric standards (Wade \&
Horne 1988; Matheson et al. 2001).  The spectra were, in general,
taken at or near the parallactic angle (Filippenko 1982).  The
relative fluxes are thus accurate to $\sim$ 5\% over the entire
wavelength range.  Figure \ref{grb-all} shows a signal-to-noise ratio
weighted average our spectra.  The initial report on these data was
presented by Castander et al. (2002).

\section{Results}

\begin{figure}
\plotone{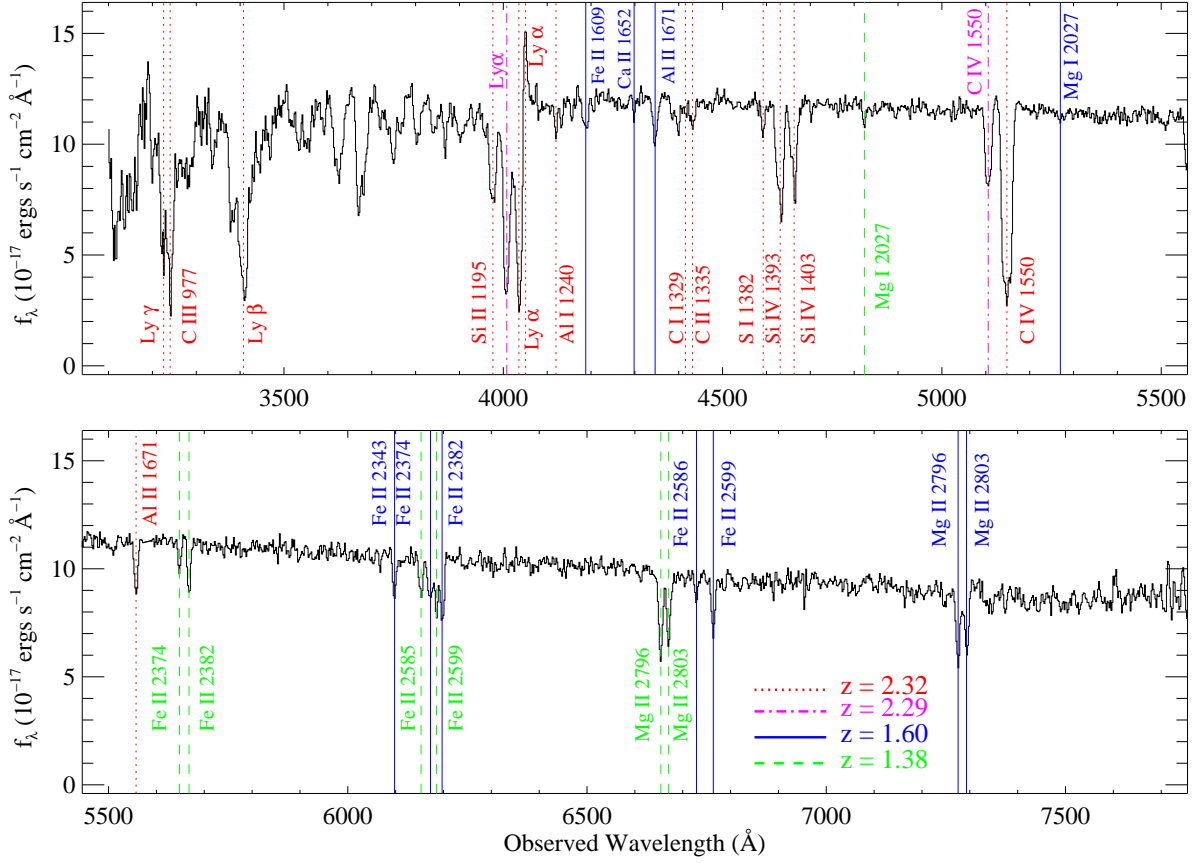}
\caption{Spectrum of the OT associated with GRB 021004, showing a blue
continuum with three distinct absorption systems.  This is the
combination of all observations, weighted by the signal-to-noise ratio
of each individual spectrum.  Prominent absorption features associated
with the three redshift systems are indicated: $z =
1.38$---\emph{dashed vertical lines}; $z = 1.60$---\emph{solid
vertical lines}; $z = 2.33$---\emph{dotted vertical lines}.  The lines
of Ly$\alpha$ and \ion{C}{4} $\lambda\lambda$1448, 1551 at $z = 2.29$
that probably represent outflows from the $z = 2.33$ host are also
marked (\emph{dash-dotted vertical lines}).  The unidentified lines
blueward of Ly$\alpha$ are most likely due to Ly$\alpha$ clouds along
the line of sight.\label{grb-all}}
\end{figure}

The spectrum of the OT of GRB 021004 (Figure \ref{grb-all}) consists
of a blue continuum typical for GRBs.  There are several absorption
features identifiable as two intervening metal-line systems at
redshifts of $z = 1.380$ and $z = 1.602$ (first reported by Fox et
al. 2002b) and one set of lines at a redshift of $z = 2.323$,
apparently intrinsic to the host galaxy of the GRB (first noted by
Chornock \& Filippenko 2002).  The three systems are marked on the
spectrum in Figure \ref{grb-all} and equivalent widths are reported in
Table 2.

The two lower-redshift \ion{Mg}{2} systems are fairly typical (e.g.,
Steidel \& Sargent 1992).  The system at $z = 2.323$ has lines with
large equivalent widths, as well as distinct absorption and
\emph{emission} from Ly$\alpha$, indicating that it most likely
represents the host galaxy of the GRB.  Our low spectral resolution
precludes identification of all separate components of the lines noted
above.  There are strong blueshifted lines apparent for \ion{C}{4}
$\lambda\lambda$1548, 1551 at a rest-frame velocity of $\sim 2400$ km
s$^{-1}$ and Ly$\alpha$ at a rest-frame velocity of $\sim 1500$ km
s$^{-1}$, although \ion{C}{4} $\lambda\lambda$1548, 1551 at $z = 1.60$
most likely contaminates the Ly$\alpha$ absorption, which may explain
the discrepancy between the two velocities.

\begin{figure}
\plotone{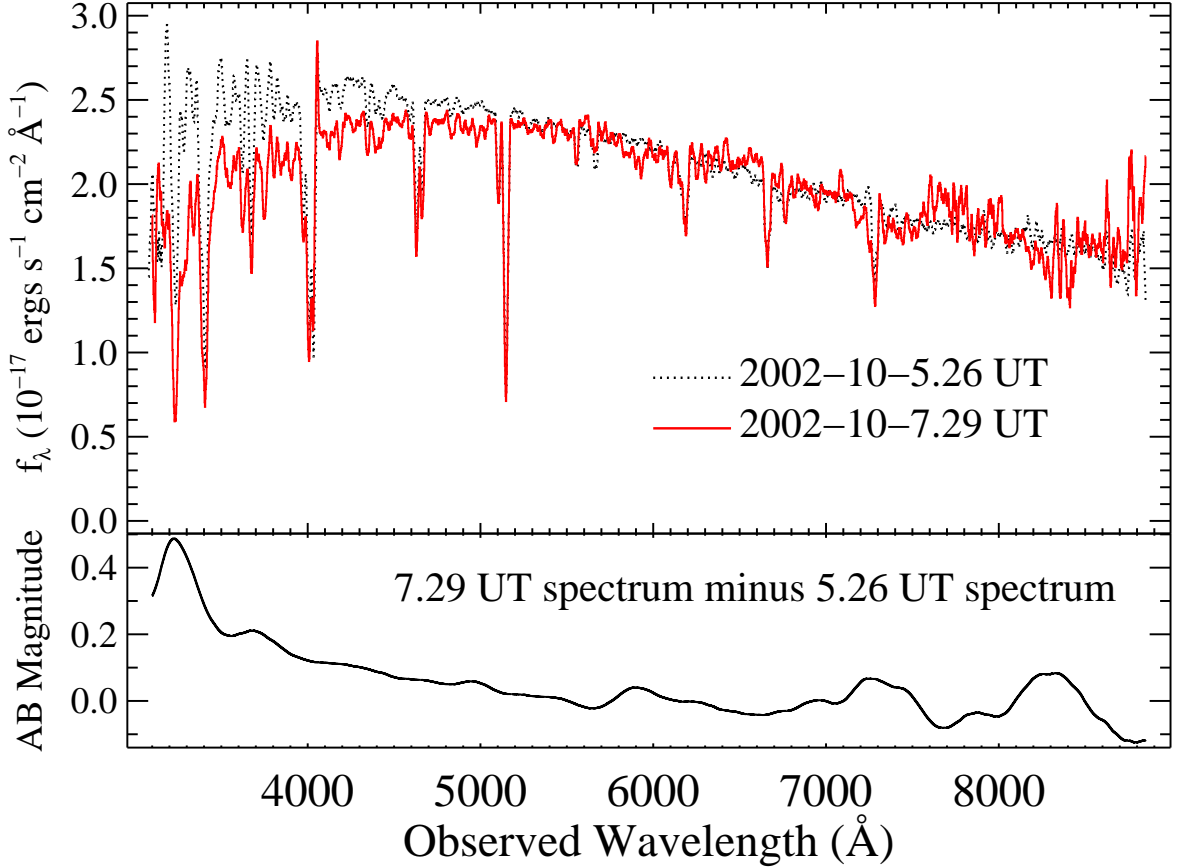}
\caption{Spectra of the OT associated with GRB 021004 at two different
epochs (\emph{top panel}), showing a distinct color change
({\emph{bottom panel}).  The top panel shows a spectrum taken at
2002-10-5.26 (sum of two exposures at 5.24 and 5.27 UT---\emph{dotted
line}) and one taken at 2002-10-7.29 UT (sum of two exposures at 7.28
and 7.30 UT--\emph{solid line}).  The spectra have been normalized
over the range 5000-7000\AA\ and smoothed by a boxcar of width 22\AA.
The 7.29 UT spectrum is fainter at the blue end, shown graphically in
the bottom panel.  This is the difference between the 7.29 UT spectrum
and the 5.26 spectrum, heavily smoothed to emphasize the continuum
shape.  The units are AB magnitudes (Oke \& Gunn 1983).  The artifacts
in the smoothed difference spectrum reflect the noise in the data, and
indicate that the error in this color curve is $\sim$ 0.1 mag.  There
is a clear difference at the blue end, implying a color change of
$\sim$ 0.3 mag from 5.26 UT to 7.29 between the approximate ranges of
the $B$ and $V$ filters.  This scale is borne out by simultaneous
photometry (Bersier et al. 2002b), and is the first clear
spectroscopic evidence for a color change in the OT of a
GRB\label{grbcolor}.}}
\end{figure}

The brightness of the OT and its relatively slow rate of decline (Winn
et al. 2002; Bersier et al. 2002a; Stanek et al. 2002) allowed us to
observe it over three days with the MMT, thus providing an opportunity
to look for spectroscopic evolution.  There was no clear change in
absorption line structure or strength from Oct 5.12 UT to Oct 7.30 UT.
There was, however, a change in continuum shape, illustrated in Figure
\ref{grbcolor}.  The comparison is made between the average of the
last two spectra from the MMT on Oct 5 (Oct 5.24 and Oct 5.27) and the
average of the two spectra from the MMT on Oct 7 (Oct 7.28 and Oct
7.30).  The spectra were normalized to match continua between 5000 and
7000\AA.  There is a distinct change in the continuum, with the Oct 7
spectrum being fainter at the blue end.  The bottom panel of Figure
\ref{grbcolor} shows the (heavily smoothed) difference between Oct
5.26 UT and Oct 7.29 UT in AB (monochromatic) magnitudes (Oke \& Gunn
1983).  Features that remain in this color curve are chiefly due to
night-sky contamination of the spectrum; they imply a conservative
estimate of the error of $\sim$ 0.1 mag.  The color curve suggests
that there should be little change in $V-R$, but that $B-V$ or $B-R$
should increase.  Our simultaneous photometry
(Bersier et al. 2002b) indicates that a change in color of
$0.2-0.3$ mag does occur over this time period.  We also have spectra
from Oct 6 UT, the intervening night.  The average of the two spectra
from that night, Oct 6.27 UT and Oct 6.34 UT, does fall in between the
spectra from Oct 5 and Oct 7.  We chose, however, not to include it on
Figure \ref{grbcolor} for the sake of clarity.

Color changes in an OT have been observed before (e.g., Galama et
al. 1998) using broad-band photometry.  While spectroscopic evidence
is more difficult to obtain, the additional information about how the
color is changing at every wavelength provides a stronger constraint
for models of the afterglow.  

The spectral index also indicates spectroscopic variation over the
three-day period of observation.  Assuming a Galactic color excess of
$E(B-V) = 0.06$ (Schlegel, Finkbeiner, \& Davis 1998), we dereddened
the spectra using the extinction correction of Cardelli, Clayton, \&
Mathis (1989), with the O'Donnell (1994) modifications at blue
wavelengths.  Least-squares minimization for a power law ($F_{\nu}
\propto \nu^{-\beta}$) over the full wavelength range
($3100-8850$\AA) yields an index of $\beta = 1.42 \pm 0.02$
(statistical error only) for Oct 5
UT, but there is clearly curvature present in the spectrum.  Fitting
only in the range $5500-8850$ gives an index of $\beta = 0.95 \pm
0.03$.  By Oct 7 UT, the power-law index for the full range of the
spectrum is $\beta = 1.67 \pm 0.02$, but only $\beta = 0.98 \pm 0.03$
for $5500-8850$, fully consistent with the value from the first night.
This again shows that the spectroscopic variation is real, but
confined to the blue wavelengths.

Slight differences at the blue end of an optical spectrum can easily
be caused by atmospheric dispersion unless the spectrograph slit is
oriented at the parallactic angle (Filippenko 1982).  All of our
observations with the MMT presented here were taken at or near the
parallactic angle, as were the exposures of the standard stars used to
calibrate the data.  Moreover, the spectra used to calculate the color
difference shown in Figure \ref{grbcolor} were all taken at an airmass
of less than 1.07, implying that the effects of atmospheric dispersion
were negligible.  As mentioned in Table 1, there is second-order light
contaminating the red end of the MMT spectra.  Given that both the
spectra from FLWO (that are not affected by second-order light) and
approximately simultaneous photometry (Bersier et al. 2002b) agree
with the shape of the MMT spectra, we believe that we have minimized
the effects of the second-order contamination.

The deviation from a simple power law evident in the first spectrum
from Oct 5 UT may be intrinsic to the OT or possibly the result of
extinction in the host galaxy.  There are strong absorption features
due to the host, so there is clearly a considerable amount of
intervening material at the host redshift.  The fact that the spectrum
changes over the three days of our observations, though, shows that
there is more than just extinction at work.  If GRBs do destroy dust
along the line of sight (Fruchter, Krolik, \& Rhoads 2001; Draine \&
Hao 2002; Perna \& Lazzati 2002), the decrease in blue flux we see
would not be the signature.  Several of the unidentified features in
the spectrum, especially blueward of Ly$\alpha$, could be the result
of the H$_2$ absorption as described by Draine \& Hao (2002), but they
are more likely the result of absorption by Ly$\alpha$ clouds.  One
possible explanation for the reddening of the spectrum would be the
formation of dust.  Rhoads, Burud, \& Fruchter (2002) reported that
the OT decreased in brightness in the $H$ band (compared to the
optical photometry of Stanek et al. 2002) over the same time
period as our spectrum shows a decrease at blue optical wavelengths.
This suggests that dust formation is unlikely.  

The increase in $(B-V)$ color seen in our data, along with the
decrease in $(R-H)$ color described by Rhoads et al. (2002), supports
the suggestion of Rhoads et al. that this could be a cooling break
affecting longer optical wavelengths.  If this scenario is correct,
our data would be most useful for constraining the shape of the
spectral break in the afterglow (Granot \& Sari 2002).

The nature of GRBs remains a powerful enigma.  Optical
spectroscopy has played a critical role in establishing that GRBs are
at cosmological distances (e.g., Metzger et al. 1997), and it has begun
 to illuminate other characteristics.  The color change
that we see represents the first conclusive evidence for variability over
time in spectroscopy. 

The MMT spectra described in this paper are available for analysis by
other researchers via anonymous ftp at
\texttt{ftp://cfa-ftp.harvard.edu/pub/kstanek/GRB021004\_spec/}.

\acknowledgments 
We would like to thank Adam Dobrzycki, Brian McLeod,
and Rosalba Perna for useful discussions.  We also thank the anonymous
referee for valuable comments.

\begin{deluxetable}{cccccccccccc}
\tabletypesize{\scriptsize}
\rotate
\tablenum{1}
\tablecaption{JOURNAL OF OBSERVATIONS}
\tablehead{\colhead{UT Date\tablenotemark{a}} &
\colhead{Julian Day\tablenotemark{b}} &
\colhead{Age\tablenotemark{c}} &
\colhead{Tel.} &
\colhead{Range}  &
\colhead{Res.\tablenotemark{d}} &
\colhead{P.A.\tablenotemark{e}} &
\colhead{Par.\tablenotemark{f}} &
\colhead{Air.\tablenotemark{g}} & 
\colhead{Flux Std.\tablenotemark{h}} &
\colhead{Slit} &
\colhead{Exp.}  \\
\colhead{} &
\colhead{} &
\colhead{(hours)} &
\colhead{} &
\colhead{(\AA)} &
\colhead{(\AA)} &
\colhead{($^\circ$)} &
\colhead{($^\circ$)} &
\colhead{} &
\colhead{} &
\colhead{($^{\prime\prime}$)} &
\colhead{(s)} }
\startdata
 2002-10-05.09 & 2552.59 & 14.2 & Baade &  4000-9000 & 13.1 & +28.8 & -38.8 & 1.82 & \nodata\tablenotemark{i} &  1.25 &  600  \\

 2002-10-05.10 & 2552.60 & 14.4 & MMT\tablenotemark{j} &  3080-8850 & 11.0 & -63.2 &  -56.99 & 2.22 & BD28/BD17 &  2 &  600  \\
 2002-10-05.10 & 2552.60 & 14.4 & Baade &  4050-6850 & 6.2 & +28.8 & -38.8 & 1.75 & \nodata\tablenotemark{i} &  1.25 &  900  \\

 2002-10-05.12 & 2552.62 & 14.9 & MMT\tablenotemark{j} &  3080-8850 & 11.0 & -63.8 &  -57.16 & 1.79 & BD28/BD17 &  2 & 1800  \\
 2002-10-05.15 & 2552.65 & 15.6 & MMT &  4900-9000 & 11.0 & -64.1 &  -56.32 & 1.48 & BD17 &  2 & 1800  \\
 2002-10-05.17 & 2552.67 & 16.1 &FLWO &3720-7540.5 & 7.0 &  -70.0 &  -63.08 & 1.32 & BD28 &  3 & 1800  \\
 2002-10-05.19 & 2552.69 & 16.6 &FLWO &3720-7540.5 & 7.0 &  -70.0 &  -61.06 & 1.21 & BD28 &  3 & 1800  \\
 2002-10-05.21 & 2552.71 & 17.0 & Baade &  4050-6850 & 6.2 & -18.7 & -80.7 & 1.53 & \nodata\tablenotemark{i} &  1.25 &  1200  \\

 2002-10-05.24 & 2552.74 & 17.8 & MMT\tablenotemark{j} &  3080-8850 & 11.0 & -52.1 &  -35.97 & 1.07 & BD28/BD17 &  2 & 1800  \\
 2002-10-05.27 & 2552.77 & 18.5 & MMT\tablenotemark{j} &  3080-8850 & 11.0 & -36.9 &  -18.66 & 1.03 & BD28/BD17 &  2 & 1800  \\
 2002-10-06.27 & 2553.77 & 42.5 & MMT\tablenotemark{j} &  3100-8850 & 11.0 & -33.5 &  -15.43 & 1.03 & BD28/BD17 &  2 & 1800  \\
 2002-10-06.34 & 2553.84 & 44.2 & MMT\tablenotemark{j} &  3100-8850 & 11.0 &  46.8 &   40.19 & 1.08 & BD28/BD17 &  2 & 1800  \\
 2002-10-07.28 & 2554.78 & 66.7 & MMT\tablenotemark{j} &  3100-8850 & 11.0 & -20.5 &   -4.63 & 1.03 & BD28/BD17 &  2 & 1800  \\
 2002-10-07.30 & 2554.80 & 67.2 & MMT\tablenotemark{j} &  3100-8850 & 11.0 &   9.4 &   14.33 & 1.03 & BD28/BD17 &  2 & 1630  \\
\enddata
\tablenotetext{a}{UT at midpoint of exposure.}
\tablenotetext{b}{Julian Date -- 2,450,000.}
\tablenotetext{c}{Age in hours from detection of the burst at UT
2002-10-04T12:06:13.57 (Shirasaki et al. 2002).}
\tablenotetext{d}{Approximate spectral resolution (full width at half
maximum intensity) as estimated from sky lines.}
\tablenotetext{e}{Position angle of the spectrograph slit.}
\tablenotetext{f}{Average parallactic angle over the course of the exposures.}
\tablenotetext{g}{Average airmass of observations.}
\tablenotetext{h}{The standard stars are as follows: BD28 =
BD+28$^\circ$4211, ---Stone (1977), Massey \& Gronwall (1990); BD17 =
BD+17$^\circ$4708---Oke \& Gunn (1983).}
\tablenotetext{i}{Problems with the Magellan data prevented accurate
 flux calibration.  These spectra were instead matched to the shape of the
 temporally closest MMT spectrum for the final, combined spectrum.}
\tablenotetext{j}{There is some second-order light contaminating the
 red wavelengths of these spectra.  Through the use of standard stars
 of different colors, we believe we have minimized any impact of this.}
\end{deluxetable}

\begin{deluxetable}{lcccc}
\tablewidth{0pt}
\tablenum{2}
\tablecaption{Absorption-Line Identifications in the Spectrum of GRB 021004\label{lineids}}
\tablehead{\colhead{Observed} &
\colhead{} &
\colhead{Rest} &
\colhead{Rest-Frame} &
\colhead{} \\
\colhead{Wavelength} &
\colhead{Line} &
\colhead{Wavelength} &
\colhead{Equivalent Width} &
\colhead{} \\
\colhead{(\AA)} &
\colhead{Identification} &
\colhead{(\AA)} &
\colhead{(\AA)} &
\colhead{Redshift}}
\startdata
7293.1 & \ion{Mg}{2} & 2802.7 & 1.2 $\pm$ 0.2 & 1.602 \\
7275.5 & \ion{Mg}{2} & 2795.5 & 1.5 $\pm$ 0.2 & 1.603 \\
6763.9 & \ion{Fe}{2} & 2599.4 & 1.0 $\pm$ 0.2 & 1.602 \\
6728.8 & \ion{Fe}{2} & 2585.9 & 0.4 $\pm$ 0.2 & 1.602 \\
6670.5 & \ion{Mg}{2} & 2802.7 & 1.3 $\pm$ 0.2 & 1.380 \\
6654.1 & \ion{Mg}{2} & 2795.5 & 1.6 $\pm$ 0.2 & 1.380 \\
6197.1 & \ion{Fe}{2} & 2382.0 & 1.1 $\pm$ 0.3 & 1.602 \\
6185.4 & \ion{Fe}{2} & 2599.4 & 0.8 $\pm$ 0.3 & 1.380 \\
6172.8 & \ion{Fe}{2} & 2373.7 & 1.0 $\pm$ 0.3 & 1.601 \\
6153.2 & \ion{Fe}{2} & 2585.9 & 0.7 $\pm$ 0.3 & 1.380 \\
6097.8 & \ion{Fe}{2} & 2343.5 & 0.6 $\pm$ 0.3 & 1.602 \\
5668.1 & \ion{Fe}{2} & 2382.0 & 0.7 $\pm$ 0.2 & 1.380 \\
5648.3 & \ion{Fe}{2} & 2373.7 & 0.4 $\pm$ 0.2 & 1.380 \\
5557.5 & \ion{Al}{2} & 1670.8 & 0.6 $\pm$ 0.2 & 2.326 \\
5270.0 & \ion{Mg}{1} & 2026.9 & 0.2 $\pm$ 0.2 & 1.600 \\
5148.3 & \ion{C}{4}  & 1550.8 & 5.7 $\pm$ 0.3 & 2.320 \\
5105.6 & \ion{C}{4}  & 1550.8 & 1.5 $\pm$ 0.3 & 2.292 \\
4823.4 & \ion{Mg}{1} & 2026.9 & 0.2 $\pm$ 0.2 & 1.380 \\
4663.0 & \ion{Si}{4}  & 1402.8 & 1.5 $\pm$ 0.2 & 2.324 \\
4631.3 & \ion{Si}{4}  & 1393.8 & 2.1 $\pm$ 0.2 & 2.323 \\
4592.5 & \ion{S}{1}  & 1381.6 & 0.3 $\pm$ 0.2 & 2.324 \\
4431.3 & \ion{C}{1}  & 1329.1 & 0.4 $\pm$ 0.3 & 2.334 \\
4346.2 & \ion{Al}{2} & 1670.8 & 0.8 $\pm$ 0.2 & 1.601 \\
4298.2 & \ion{Ca}{2} & 1652.0 & 0.1 $\pm$ 0.2 & 1.602 \\
4187.9 & \ion{Fe}{2} & 1608.5 & 0.6 $\pm$ 0.4 & 1.604 \\
4120.1 & \ion{Al}{1} & 1240.4 & 0.2 $\pm$ 0.3 & 2.322 \\
4050.8 & Ly $\alpha$\tablenotemark{a} & 1215.7 & 0.6 $\pm$ 0.3 & 2.332 \\
4035.6 & Ly $\alpha$ & 1215.7 & 3.6 $\pm$ 0.3 & 2.320 \\
4008.0 & Ly $\alpha$ & 1215.7 & 4.2 $\pm$ 0.4 & 2.297 \\
3976.3 & \ion{Si}{2}  & 1194.5 & 2.0 $\pm$ 0.4 & 2.329 \\
3407.7 & Ly $\beta$ & 1025.7 & 6.0 $\pm$ 0.5 & 2.322 \\
3225.6 & Ly $\gamma$ & 972.54 & 3.9 $\pm$ 0.5 & 2.317 \\
\enddata 
\tablenotetext{a}{Emission component.}
\end{deluxetable} 
\clearpage

\end{document}